\documentclass[11pt]{article}
\usepackage[centertags]{amsmath}
\usepackage{amssymb}
\usepackage{graphicx}

\begin{document}

\title{\bf Waiting Cycle Times and Generalized Haldane Equality in the Steady-state Cycle Kinetics of Single Enzymes}

\author{Hao Ge\footnote{
School of Mathematical Sciences, Peking University, Beijing 100871,
P.R.China; email: edmund\_ge@tom.com}}

\maketitle{}
\date{}

\begin{abstract}
Enzyme kinetics are cyclic. A more realistic reversible three-step
mechanism of the Michaelis-Menten kinetics is investigated in
detail, and three kinds of waiting cycle times $T$, $T_{+}$, $T_{-}$
are defined. It is shown that the mean waiting cycle times $\langle
T \rangle$, $\langle T_{+} \rangle$, and $\langle T_{-} \rangle$ are
the reciprocal of the steady-state cycle flux $J^{ss}$, the forward
steady-state cycle flux $J^{ss}_{+}$ and the backward steady-state
cycle flux $J^{ss}_{-}$ respectively. We also show that the
distribution of $T_{+}$ conditioned on $T_{+}<T_{-}$ is identical to
the distribution of $T_{-}$ conditioned on $T_{-}<T_{+}$, which is
referred as generalized Haldane equality. Consequently, the mean
waiting cycle time of $T_{+}$ conditioned on $T_{+}<T_{-}$ ($\langle
T_{+}| T_{+}<T_{-} \rangle$) and the one of $T_{-}$ conditioned on
$T_{-}<T_{+}$ ($\langle T_{-}| T_{-}<T_{+} \rangle$) are both just
the same as $\langle T \rangle$. In addition, the forward and
backward stepping probabilities $p^{+},p^{-}$ are also defined and
discussed, especially their relationship with the cycle fluxes and
waiting cycle times. Furthermore, we extend the same results to the
$n$-step cycle, and finally, experimental and theoretically based
evidences are also included.
\begin{flushleft}
\end{flushleft}
\end{abstract}

\section{Introduction}

Living cells function thermodynamically as open systems that are far
from static thermal equilibrium, since cells must continually
extract energy from their surroundings in order to sustain the
characteristic features of life such as growth, cell division,
intercellular communication, movement and responsiveness to their
environment.

From the view of statistical physics, these stochastic models for
systems biology exhibit nonequilibrium steady states (NESS) in which
nonequilibrium circulations (cycle fluxes) necessarily emerge
 \cite{hqjpcm05}. Hong Qian and his co-workers have recently
discussed the relation between an NESS and traditional nonlinearly
dynamics \cite{QE,hqprl05,hqjpc06,QH1}.

The researches on irreversible systems far from equilibrium began
with the works by Haken  \cite{Hak77,Hak83} about laser and
Prigogine, etc.  \cite{GlPr,NP} about oscillations of chemical
reactions. It is closely related to another concept of macroscopic
irreversibility in nonequilibrium statistical physics. A macroscopic
irreversible system in a steady state should have positive entropy
production rate and should be in nonequilibrium.

T.L. Hill, etc.  \cite{Hi66,Hi77,Hi95,HC} constructed a general
mesoscopic model for the combination and transformation of
biochemical polymers in vivid metabolic systems since 1966. Their
results can be applied to explain the mechanism of muscle
contraction and active transports \cite{FMWT}.

Mathematical theory of nonequilibrium steady states and circulation
(cycle fluxes) has been discussed for several decades since the
original work  \cite{QQ82,QQ85,QQG,QQQ84}, in which Qian and
co-workers developed the formulae for entropy production rate and
circulation distribution of homogeneous Markov chains, Q-processes
and diffusions, and moreover their relationship with reversibility.
They concluded that the chain or process is reversible if and only
if its entropy production vanishes, or iff there is no net cycle
fluxes. Here, we recommend a recent book  \cite{JQQ} for the
systematic presentation of this theory.

Recently, we investigate the synchronized stochastic dynamics of a
network model of yeast cell-cycle regulation  \cite{Ge4}, applying
the mathematical theory of cycle fluxes (circulation) of Markov
chains. In our model of yeast cell cycle, the trajectory
concentrates around a main cycle with the dominant circulation,
which we call  stochastic limit cycle, that is the natural
generalization of the deterministic limit cycle in the stochastic
system.

Recent advances in single-molecule spectroscopy and manipulation
have now made it possible to study enzyme kinetics at the level of
single molecules, where the stochastic effects, termed as ``dynamic
disorder'', are significant. Experimentalists can not only directly
measure the distributions of molecular properties through
single-molecule experiments rather than the ensemble average, but
also apply the theory of stochastic processes to analyze the
statistical properties of the stochastic trajectory
\cite{LXX,Xie-rev1,Xie-rev2,Xie-rev3}.

Xie, et.al \cite{Xie1,Xie2,Xie3,KCMEX} observed that the mean
waiting time is the same as the reciprocal of the Michaelis-Menten
steady-state flux (i.e., the cycle flux in my language). But the
model they built in their theoretical analysis is the simplest
irreversible Michaelis-Menten mechanism, and the state space of
their stochastic model (Markov chain) actually only contains two
states ($E$ and $ES$), which does not distinguish the two different
pathways from $ES$ to $E$ and is always in mathematical detailed
balance rather than chemical detailed balance. That is just why they
can only directly using the ordinary differential equations to get
the explicit distribution function $f(t)$ of the waiting time, and
avoid applying the strong Markov property, which is the basic method
to compute mean waiting times in stochastic processes. So their
method can not be generalized to more complicated cases, and
generally speaking, the explicit distribution function $f(t)$ can
rarely be obtained in such an analytic form.

In the present paper, a more realistic reversible three-step
mechanism of the Michaelis-Menten kinetics is investigated in
detail, and three kinds of waiting cycle times $T$, $T_{+}$, $T_{-}$
are defined. It is shown that the mean waiting cycle times $\langle
T \rangle$, $\langle T_{+} \rangle$, and $\langle T_{-} \rangle$ are
the reciprocal of the steady-state cycle flux $J^{ss}$, the forward
steady-state cycle flux $J^{ss}_{+}$ and the backward steady-state
cycle flux $J^{ss}_{-}$ respectively.

We also show that the distribution of $T_{+}$ conditioned on
$T_{+}<T_{-}$ is identical to the distribution of $T_{-}$
conditioned on $T_{-}<T_{+}$, which is referred as generalized
Haldane equality \cite{QX}. This is a key result of this work. There
is experimental evidence for it, as well as theoretical models
proving equal mean time \cite{CC,KSP,KCMEX}.

Consequently, the mean waiting cycle time of $T_{+}$ conditioned on
$T_{+}<T_{-}$ ($\langle T_{+}| T_{+}<T_{-} \rangle$) and the one of
$T_{-}$ conditioned on $T_{-}<T_{+}$ ($\langle T_{-}| T_{-}<T_{+}
\rangle$) are both just the same as $\langle T \rangle$. In
addition, the forward and backward stepping probabilities
$p^{+},p^{-}$ are also defined and discussed, especially their
relationship with the cycle fluxes and waiting cycle times.
Furthermore, we extend the same results to the $n$-step cycle, and
finally, experimental and theoretically based evidences are also
included.

\section{Single enzyme kinetics: cycle flux and NESS}

This subsection is just a brief introduction to our model and cycle
fluxes, which is from Ref.  \cite{hqjpcm05}.

\begin{figure}[h]
\centerline{\includegraphics[width=3in,height=4in,angle=270]{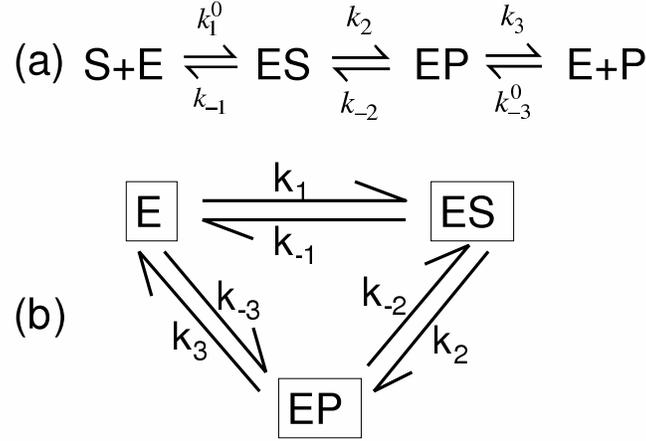}}
\caption[fig1]{Kinetic scheme of a simple reversible enzyme reaction
(a) in which $k_1^0$ and $k^0_{-3}$ are second-order rate constants.
From the perspective of a single enzyme molecule, the reaction is
unimolecular and cyclic (b). The pseudo-first-order rate constants
$k_1=k_1^0c_S$ and $k_{-3}=k_{-3}^0c_P$ where $c_S$ and $c_P$ are
the concentrations of substrate $S$ and $P$ in the steady state.}
\label{fig1}
\end{figure}

We consider a more realistic three-step mechanism of the
Michaelis-Menten kinetics in which the conversion of $S$ into $P$ in
the catalytic site of the enzyme is represented as a process
separate from release of $P$ from the enzyme (Fig. \ref{fig1}(a)):

\begin{equation}\label{Mech}
E+S\overset{k_1^0}{\underset{k_{-1}}{\rightleftharpoons}}ES\overset{k_2}{\underset{k_{-2}}{\rightleftharpoons}}EP\overset{k_3}{\underset{k_{-3}^0}{\rightleftharpoons}}E+P.
\end{equation}

If there is only one enzyme molecule, then from the enzyme
perspective, the kinetics are stochastic and cyclic, as shown in
Fig. \ref{fig1} (b), with the pseudo-first-order rate constants
$k_1=k_1^0c_S$ and $k_{-3}=k_{-3}^0c_P$ where $c_S$ and $c_P$ are
the sustained concentrations of substrate $S$ and $P$ in the steady
state.

At the chemical equilibrium, the concentrations of $S$ and $P$
satisfy $\frac{c_P}{c_S}=\frac{k_1^0k_2k_3}{k_{-1}k_{-2}k_{-3}^0}$,
i.e.

\begin{equation}
\frac{k_1k_2k_3}{k_{-1}k_{-2}k_{-3}}=1.
\end{equation}

This is the ``thermodynamic box'' in elementary chemistry, also
known as Wegscheider's relation and detailed balance. However, if
the $c_S$ and $c_P$ are maintained at constant levels that are not
at chemical equilibrium, as metabolite concentrations are in living
cells, the the enzyme reaction is in an open system that approaches
a NESS. This is the scenario in enzyme kinetics.

In this case,
\begin{equation}
\frac{k_1k_2k_3}{k_{-1}k_{-2}k_{-3}}=\gamma\neq 1,
\end{equation}
and $\triangle\mu=k_BT\ln \gamma$ is well known as the cellular
phosphorylation potential.

From the perspective of single enzyme molecule, the rate equation
for the probabilities of the states is a master equation

\begin{eqnarray}
\frac{dP_E(t)}{dt}&=&-(k_1+k_{-3})P_E(t)+k_{-1}P_{ES}(t)+k_3P_{EP}(t)\nonumber\\
\frac{dP_{ES}(t)}{dt}&=&k_1P_E(t)-(k_{-1}+k_2)P_{ES}(t)+k_{-2}P_{EP}(t)\nonumber\\
\frac{dP_{EP}(t)}{dt}&=&k_{-3}P_E(t)+k_2P_{ES}(t)-(k_{-2}+k_3)P_{EP}(t)
\end{eqnarray}

The steady-state probabilities for states $E$, $ES$ and $EP$ are
easy to compute by setting the time derivative to zero and noting
that $P_E+P_{ES}+P_{EP}=1$ for the total probability.

\begin{eqnarray}
P_E^{ss}&=&\frac{k_2k_3+k_{-1}k_3+k_{-1}k_{-2}}{k_1k_2+k_2k_3+k_3k_1+k_{-1}k_{-3}+k_{-2}k_{-3}+k_{-1}k_{-2}+k_1k_{-2}+k_2k_{-3}+k_3k_{-1}},
\nonumber\\
P_{ES}^{ss}&=&\frac{k_1k_3+k_{-2}k_{-3}+k_1k_{-2}}{k_1k_2+k_2k_3+k_3k_1+k_{-1}k_{-3}+k_{-2}k_{-3}+k_{-1}k_{-2}+k_1k_{-2}+k_2k_{-3}+k_3k_{-1}},
\nonumber\\
P_{EP}^{ss}&=&\frac{k_1k_2+k_2k_{-3}+k_{-1}k_{-3}}{k_1k_2+k_2k_3+k_3k_1+k_{-1}k_{-3}+k_{-2}k_{-3}+k_{-1}k_{-2}+k_1k_{-2}+k_2k_{-3}+k_3k_{-1}}.\nonumber\\
\end{eqnarray}

Then, the clockwise steady-state cycle flux in Fig. \ref{fig1}(b),
which is precisely the enzyme turnover rate of $S\rightarrow P$ in
Fig. \ref{fig1}(a),
$J^{ss}=P_E^{ss}k_1-P_{ES}^{ss}k_{-1}=P_{ES}^{ss}k_2-P_{EP}^{ss}k_{-2}=P_{EP}^{ss}k_3-P_{E}^{ss}k_{-3}$,
which follows

\begin{equation}
J^{ss}=\frac{k_1k_2k_3-k_{-1}k_{-2}k_{-3}}{k_1k_2+k_2k_3+k_3k_1+k_{-1}k_{-3}+k_{-2}k_{-3}+k_{-1}k_{-2}+k_1k_{-2}+k_2k_{-3}+k_3k_{-1}}=J^{ss}_{+}-J^{ss}_{-},
\end{equation}
where
$$J^{ss}_{+}=\frac{k_1k_2k_3}{k_1k_2+k_2k_3+k_3k_1+k_{-1}k_{-3}+k_{-2}k_{-3}+k_{-1}k_{-2}+k_1k_{-2}+k_2k_{-3}+k_3k_{-1}},$$
is the forward cycle flux, and
$$J^{ss}_{-}=\frac{k_{-1}k_{-2}k_{-3}}{k_1k_2+k_2k_3+k_3k_1+k_{-1}k_{-3}+k_{-2}k_{-3}+k_{-1}k_{-2}+k_1k_{-2}+k_2k_{-3}+k_3k_{-1}},$$
is the backward cycle flux.

The net cycle flux is just the Michaelis-Menten steady-state flux of
(\ref{Mech}), i.e.
$$v=\frac{k_Sc_S-k_Pc_P}{1+\frac{c_S}{K_{mS}}+\frac{c_P}{K_{mP}}},$$
where $k_S=\frac{k_1^0k_2k_3}{k_{-1}k_{-2}+k_{-1}k_3+k_2k_3}$,
$k_P=\frac{k_{-1}k_{-2}k_{-3}^0}{k_{-1}k_{-2}+k_{-1}k_3+k_2k_3}$,
$K_{mS}=\frac{k_{-1}k_{-2}+k_{-1}k_3+k_2k_3}{k_1^0(k_{-2}+k_2+k_3)}$,
and
$K_{mP}=\frac{k_{-1}k_{-2}+k_{-1}k_3+k_2k_3}{(k_{-2}+k_2+k_{-1})k_{-3}^0}$.
That is just Eq. (2.46) in  \cite{CB}.

 In addition, $J^{ss}_{+}$ and $J^{ss}_{-}$ can be rigorously
proved to be the averaged numbers of the forward and backward cycles
per time respectively due to ergodic theory \cite[Theorem
2.1.2]{JQQ}, i.e.

\begin{eqnarray}\label{erg-cycle1}
J^{ss}&=&\lim_{t\rightarrow\infty} \frac{1}{t} \nu(t)
,\nonumber\\
J_{+}^{ss}&=&\lim_{t\rightarrow\infty} \frac{1}{t} \nu_{+}(t)
,\nonumber\\
J_{-}^{ss}&=&\lim_{t\rightarrow\infty} \frac{1}{t} \nu_{-}(t),
\end{eqnarray}
where $\nu_{+}(t)$ and $\nu_{-}(t)$ are the number of occurrences of
forward and backward cycles up to time $t$, and
$\nu(t)=\nu_{+}(t)-\nu_{-}(t)$.

At the end of this section, it is important to notice that the
quantity $\gamma$ can be approximated by
$\frac{\nu_{+}(t)}{\nu_{-}(t)}$ in single-molecule experiment when
the time $t$ is large enough, due to the fact that
$\gamma=\frac{J_{+}^{ss}}{J_{-}^{ss}}$ and $J_{+}^{ss}=J_{-}^{ss}$
(i.e. $\gamma=1$) if and only if this system is at chemical
equilibrium.

\section{Waiting cycle times and generalized Haldane equality}

\subsection{Mean waiting cycle times}

The most obvious feature of the turnover trajectory (Fig. 1B in
\cite{LXX}) is its stochastic nature, exhibited both in the time
needed for a chemical reaction which takes place on the
subpicosecond time scale and that needed for diffusion and thermal
activation which is much more longer. In the single-molecule
experiment \cite{LXX}, the emission on-time and off-time recorded
correspond to the ``waiting time'' for the turnover reactions,
respectively. Once holding the statistical data of the trajectory in
hand, the most straightforward analysis of the trajectories is
certainly the distribution of the on-and-off times, so in our
theoretical model, waiting cycle times should be defined and their
mean should also be calculated at the first step.

Starting from the free enzyme state $E$, three kinds of waiting
cycle times can be defined.  $T$ represents the waiting time for the
occurrence of a forward or a backward cycle,  $T_{+}$ represents the
waiting time for the occurrence of a forward cycle, and  $T_{-}$
represents the waiting time for the occurrence of a backward cycle
respectively. Obviously, $T$ is just the smaller one of $T_{+}$ and
$T_{-}$.

\begin{figure}[h]
\centerline{\includegraphics[width=3in,height=4in,angle=270]{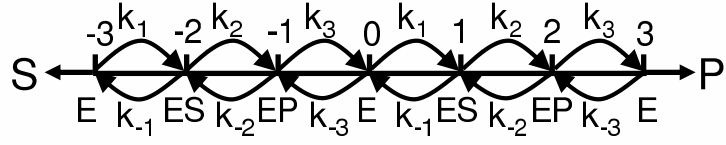}}
\caption[fig2]{The kinetic scheme for computing the waiting cycle
times. In order to distinguish the forward and backward cycles, Fig.
\ref{fig1} is transferred into a one-dimensional random walk model.}
\label{fig2}
\end{figure}

The problem of computing the mean waiting time $\langle T \rangle$
can be transferred into an important application of
first-passage-time(FPT) methods (Fig.\ref{fig2}) to the cyclic
chemical transformations, in particular single-enzyme kinetics
(Fig.\ref{fig1}(b)).

FPT problems are been well studied, and there are analytical results
for first-passage times in a discrete-time one-dimensional
asymmetric random walk for quenched disorder \cite{PC}. But actually
what we investigated in the present paper is the continuous-time
case rather than the discrete-time case in \cite{PC}, and the master
equation (\ref{Mean-cycletime}) below is different from
\cite[Eq.(2.6)]{PC}. Therefore, the expression of $\langle T
\rangle$ cannot be regarded as a particular case of the more general
FPT problem when $M=L=2$ in \cite[Eq.(2.11)]{PC} although they are
quite similar.

Let $\tau_i$ be the mean  time first hitting the state $3$ or $-3$
in Fig.\ref{fig2}, starting from the state $i$. Obviously, $\langle
T \rangle=\tau_0$ and $\tau_3=\tau_{-3}=0$.

Applying the strong Markov property of continuous-time Markov chains
\cite{And}, $\{\tau_i\}$ satisfies the following equations

\begin{eqnarray}\label{Mean-cycletime}
\tau_{-2}&=&\frac{1}{k_{-1}+k_2}+\frac{k_{-1}}{k_{-1}+k_2}\times
0+\frac{k_2}{k_{-1}+k_2}\tau_{-1},\nonumber\\
\tau_{-1}&=&\frac{1}{k_{-2}+k_3}+\frac{k_{-2}}{k_{-2}+k_3}
\tau_{-2}+\frac{k_3}{k_{-2}+k_3}\tau_0,\nonumber\\
\tau_0&=&\frac{1}{k_{-3}+k_1}+\frac{k_{-3}}{k_{-3}+k_1}
\tau_{-1}+\frac{k_1}{k_{-3}+k_1}\tau_1,\nonumber\\
\tau_1&=&\frac{1}{k_{-1}+k_2}+\frac{k_{-1}}{k_{-1}+k_2}
\tau_0+\frac{k_2}{k_{-1}+k_2}\tau_2,\nonumber\\
\tau_2&=&\frac{1}{k_{-2}+k_3}+\frac{k_{-2}}{k_{-2}+k_3}
\tau_1+\frac{k_3}{k_{-2}+k_3}\times 0.
\end{eqnarray}

Through simple calculation, one can get that
$$\langle T \rangle=\frac{k_1k_2+k_2k_3+k_3k_1+k_{-1}k_{-3}+k_{-2}k_{-3}+k_{-1}k_{-2}+k_1k_{-2}+k_2k_{-3}+k_3k_{-1}}{k_1k_2k_3+k_{-1}k_{-2}k_{-3}}
=\frac{1}{J^{ss}_{+}+J^{ss}_{-}}.$$

Similarly, another mean waiting cycle time $\langle T_{+} \rangle$,
which is the mean time to complete the forward cycle in
Fig.\ref{fig1}(b) whether before or after reaching an ``analogous''
final state in the opposite direction, can also be obtained as
solutions of the nearly identical equations to
(\ref{Mean-cycletime}) but with modified boundary conditions. Let
$\tau_{i+}$ be the mean time first hitting the state $3$, whether
before or after the time hitting the state $-3$ in Fig.\ref{fig2},
starting from the state $i$. Obviously, $\langle T_{+}
\rangle=\tau_{0+}$, $\tau_{3+}=0$ and $\tau_{-3+}=\tau_{0+}$.

Applying the strong Markov property of Markov chains again,
$\{\tau_{i+}\}$ satisfies the following equations

\begin{eqnarray}
\tau_{-2+}&=&\frac{1}{k_{-1}+k_2}+\frac{k_{-1}}{k_{-1}+k_2}\tau_{-3+}+\frac{k_2}{k_{-1}+k_2}\tau_{-1+},\nonumber\\
\tau_{-1+}&=&\frac{1}{k_{-2}+k_3}+\frac{k_{-2}}{k_{-2}+k_3}
\tau_{-2+}+\frac{k_3}{k_{-2}+k_3}\tau_{0+},\nonumber\\
\tau_{0+}&=&\frac{1}{k_{-3}+k_1}+\frac{k_{-3}}{k_{-3}+k_1}
\tau_{-1+}+\frac{k_1}{k_{-3}+k_1}\tau_{1+},\nonumber\\
\tau_{1+}&=&\frac{1}{k_{-1}+k_2}+\frac{k_{-1}}{k_{-1}+k_2}
\tau_{0+}+\frac{k_2}{k_{-1}+k_2}\tau_{2+},\nonumber\\
\tau_{2+}&=&\frac{1}{k_{-2}+k_3}+\frac{k_{-2}}{k_{-2}+k_3}
\tau_{1+}+\frac{k_3}{k_{-2}+k_3}\times 0,
\end{eqnarray}
which gives that
$$\langle T_{+} \rangle=\frac{k_1k_2+k_2k_3+k_3k_1+k_{-1}k_{-3}+k_{-2}k_{-3}+k_{-1}k_{-2}+k_1k_{-2}+k_2k_{-3}+k_3k_{-1}}{k_1k_2k_3}
=\frac{1}{J^{ss}_{+}}.$$





Almost the same derivations can be achieved for $\langle T_{-}
\rangle$, which is the mean time to complete the backward cycle in
Fig.\ref{fig1}(b), whether before or after reaching an ``analogous''
final state in the opposite direction, immediately follows
$$\langle T_{-} \rangle=\frac{k_1k_2+k_2k_3+k_3k_1+k_{-1}k_{-3}+k_{-2}k_{-3}+k_{-1}k_{-2}+k_1k_{-2}+k_2k_{-3}+k_3k_{-1}}{k_{-1}k_{-2}k_{-3}}
=\frac{1}{J^{ss}_{-}}.$$

Surely, the expression of $\langle T_{-} \rangle$ can be directly
derived due to the symmetry of the random walk in Fig. \ref{fig2}.

The quantitative relationship between the mean waiting cycle times
($\langle T \rangle$, $\langle T_{+} \rangle$, and $\langle T_{-}
\rangle$) and the cycle fluxes ($J^{ss}$, $J^{ss}_{+}$, and
$J^{ss}_{-}$) in this subsection is the first chief result of the
present paper.

Consequently, $\langle T_{+} \rangle=\langle T_{-} \rangle$ if and
only if this system is at chemical equilibrium, because of
$\gamma=\frac{\langle T_{-} \rangle}{\langle T_{+} \rangle}$.
Therefore, $\gamma$ can also be measured by the ratio of averaged
forward and backward waiting cycle times up to time $t$ in the
single-molecule experiment, which is different from the measure
method introduced at the end of the previous subsection.
Nonetheless, applying the elementary renewal theorem \cite[Sec.3.4,
Theorem 4.1,4.2]{Dur}, the two methods are asymptotically the same
because $\langle T_{+} \rangle\approx \frac{t}{\nu_{+}(t)}$ and
$\langle T_{-} \rangle\approx \frac{t}{\nu_{-}(t)}$ when $t$ is
large.

\subsection{Stepping probability}

The stepping probabilities $p^{+}(t)$ and $p^{-}(t)$ up to time $t$
are just the fractions of $\nu_{+}(t)$ and $\nu_{-}(t)$,
representing the weights of the forward and backward cycles
respectively from the statistical point of view in experiments, i.e.

$$p^{+}(t)=\frac{\nu_{+}(t)}{\nu_{+}(t)+\nu_{-}(t)},~~p^{-}(t)=\frac{\nu_{-}(t)}{\nu_{+}(t)+\nu_{-}(t)}.$$

According to Eq.\ref{erg-cycle1}, one can get the eventual stepping
probability
\begin{eqnarray}
p^{+}\stackrel{def}{=}\lim_{t\rightarrow\infty}p^{+}(t)=\frac{J^{ss}_{+}}{J^{ss}_{+}+J^{ss}_{-}}=\frac{k_1k_2k_3}{k_1k_2k_3+k_{-1}k_{-2}k_{-3}},\nonumber\\
p^{-}\stackrel{def}{=}\lim_{t\rightarrow\infty}p^{-}(t)=\frac{J^{ss}_{-}}{J^{ss}_{+}+J^{ss}_{-}}=\frac{k_{-1}k_{-2}k_{-3}}{k_1k_2k_3+k_{-1}k_{-2}k_{-3}}.
\end{eqnarray}

It is necessary to point out that the stepping probabilities
$p^{+}(t)$ and $p^{-}(t)$ are random variables depending on the
trajectories, while their fluctuations tend to vanish when $t$ tends
to infinity. Hence the eventual stepping probability $p^{+}$ and
$p^{-}$ are independent with the trajectories due to the ergodic
theory.

Interesting, the forward stepping probability can also be defined as
$p^{+}\stackrel{def}{=}P_{\{E\}}(T_{+}<T_{-})$, which means the
probability that the particle first completes a forward cycle before
a backward one starting from the initial free enzyme $E$. Similarly,
the backward stepping probability can be defined as
$p^{-}\stackrel{def}{=}P_{\{E\}}(T_{-}<T_{+})$. This is the second
chief result of the present article.

This equivalence can be explicitly seen through translating this
problem to a corresponding one of the random walk in Fig.
\ref{fig2}, either.

Let $p_{i+}$ be the probability of hitting the state $3$ before $-3$
in Fig.\ref{fig2}, starting from the state $i$. Obviously,
$p_{3+}=1$ and $p_{-3+}=0$.

Again applying the strong Markov property of Markov chains as what
we have done in the precious section, $\{p_{i+}\}$ satisfies the
following equations

\begin{eqnarray}
p_{-2+}&=&\frac{k_{-1}}{k_{-1}+k_2}\times 0+\frac{k_2}{k_{-1}+k_2}p_{-1+},\nonumber\\
p_{-1+}&=&\frac{k_{-2}}{k_{-2}+k_3}
p_{-2+}+\frac{k_3}{k_{-2}+k_3}p_{0+},\nonumber\\
p_{0+}&=&\frac{k_{-3}}{k_{-3}+k_1}
p_{-1+}+\frac{k_1}{k_{-3}+k_1}p_{1+},\nonumber\\
p_{1+}&=&\frac{k_{-1}}{k_{-1}+k_2}
p_{0+}+\frac{k_2}{k_{-1}+k_2}p_{2+},\nonumber\\
p_{2+}&=&\frac{k_{-2}}{k_{-2}+k_3}
p_{1+}+\frac{k_3}{k_{-2}+k_3}\times 1.\nonumber
\end{eqnarray}

Through simple calculation, one can get that
$$p^{+}=P_{\{E\}}(T_{+}<T_{-})=p_{0+}=\frac{k_1k_2k_3}{k_1k_2k_3+k_{-1}k_{-2}k_{-3}},$$
and
$$p^{-}=P_{\{E\}}(T_{+}>T_{-})=1-P_{\{E\}}(T_{+}<T_{-})=\frac{k_{-1}k_{-2}k_{-3}}{k_1k_2k_3+k_{-1}k_{-2}k_{-3}}.$$

Consequently,
$$p^{+}=\frac{J^{ss}_{+}}{J^{ss}_{+}+J^{ss}_{-}}=\frac{\langle T \rangle}{\langle T_{+} \rangle},$$
$$p^{-}=\frac{J^{ss}_{-}}{J^{ss}_{+}+J^{ss}_{-}}=\frac{\langle T \rangle}{\langle T_{-} \rangle},$$
and
$$\triangle\mu=k_BT\log\gamma=k_BT\log\frac{p^{+}}{p^{-}}=k_BT\log\frac{J^{ss}_{+}}{J^{ss}_{-}}=k_BT\log\frac{\langle T_{-} \rangle}{\langle T_{+} \rangle},$$
which follows $p^{+}=p^{-}$ if and only if this system is at
chemical equilibrium.

\subsection{Generalized Haldane equality}

To avoid the unnecessary difficult mathematical details, we apply a
simple trick like the ``time-reversal mapping'' always used in
modern statistical physics \cite{Cro1,Cro2,Cro3,Jar1,Jar2,Jar3,Jar4}
instead of the rigorous language of measure theory.

We introduce a one-to-one mapping $r$ for the trajectory of the
simple kinetic in Fig. \ref{fig1}, which belongs to the event
$\{T_{+}<T_{-}\}$, mapped to its ``quasi-time-reversal'' one.

For each trajectory $\omega=\{\omega_t: t\geq 0,~\omega_0=\{E\}\}$
belonging to the set $\{T_{+}<T_{-}\}$, let $T^*$ be the \emph{last
time} when it leaves the state $\{E\}$ before finishing a forward
cycle in the Fig.\ref{fig1}(b). Then its ``quasi-time-reversal'' one
$r\omega=\{(r\omega)_t: t\geq 0\}$ is defined as follows:\\
i) when the time $t$ is before or equal to $T^*$, then one just copy $\omega$ to $r\omega$, i.e.$(r\omega)_t=\omega_t$;\\
ii) when the time $t$ is between $T^*$ and $T_{+}$, then one maps
the real time-reversal trajectory of $\omega$ with respect to the
time interval $[T^*,T_{+}]$ to $r\omega$, i.e.
$(r\omega)_t=\omega_{T^*+T_{+}-t}$; \\
iii) when the time $t$ is greater than $T_{+}$, then one can also
simply copy $\omega$ to $r\omega$ as what we have done in (i).

\begin{figure}[h]
\centerline{\includegraphics[width=3in,height=4in,angle=270]{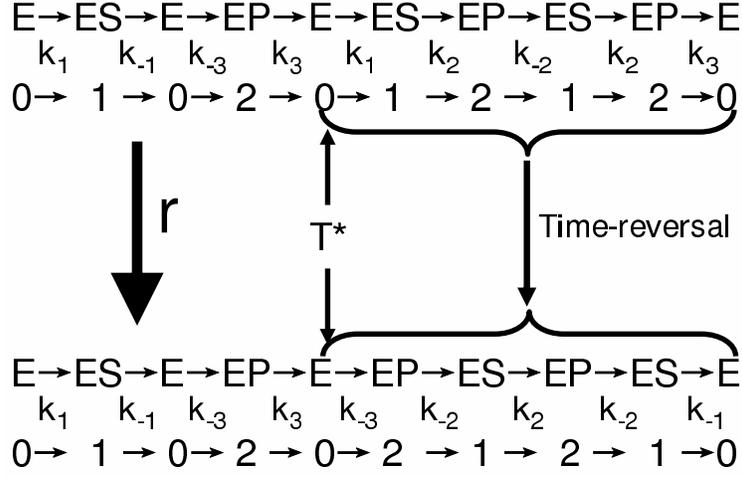}}
\caption[fig3]{An illustrative example of the
``quasi-time-reversal'' map. $T^*$ is the \emph{last time} when it
leaves the state $\{E\}$ before finishing a forward cycle
$E\rightarrow ES\rightarrow EP\rightarrow E$, then one maps the real
time-reversal trajectory of $\omega$ with respect to the time
interval $[T^*,T_{+}]$ to $r\omega$. See text for details.}
\label{fig3}
\end{figure}

See Fig. \ref{fig3} for an illustrative example. As having been
pointed out on this figure, $T^*$ is denoted to be the \emph{last
time} when it leaves the state $\{E\}$ before finishing a forward
cycle $E\rightarrow ES\rightarrow EP\rightarrow E$. Then \emph{the
ratio} of the probability density of the above trajectory with
respect to its ``quasi-time-reversal'' one below is
$$\gamma=\frac{(k_1k_{-1}k_{-3}k_3)\times(k_1k_2k_{-2}k_2k_3)}{(k_1k_{-1}k_{-3}k_3)\times(k_{-3}k_{-2}k_2k_{-2}k_{-1})}=\frac{k_1k_2k_3}{k_{-1}k_{-2}k_{-3}}.$$

Now it is indispensable to explain why we construct the above
mapping like this.

1) The number of the steps $E\rightarrow ES$ in the original
trajectory $\omega$ belonging to $\{T_{+}<T_{-}\}$ is one more than
that in its ``quasi-time-reversal'' corresponding trajectory
$r\omega$ belonging to $\{T_{+}>T_{-}\}$, while the number of the
steps $ES\rightarrow E$ in $\omega$ is one less than that in
$r\omega$;

similarly,

2) The number of the steps $ES\rightarrow EP$ in the trajectory
$\omega$ is one more than that in $r\omega$, while the number of the
steps $EP\rightarrow ES$ in the trajectory $\omega$ is one less than
that in $r\omega$;

3) The number of the steps $EP\rightarrow E$ in the trajectory
$\omega$ is one more than that in $r\omega$, while the number of the
steps $E\rightarrow EP$ in the trajectory $\omega$ is one less than
that in $r\omega$;

and more important

4) The dwell time upon each state of the trajectory $\omega$ and its
``quasi-time-reversal'' corresponding one $r\omega$ is mapped quite
well such that the difference between $\omega$ and $r\omega$ are
only exhibited upon their sequences of states.

Consequently, the most important observation is that \emph{the
ratio} of the probability density of each trajectory $\omega$ in
$\{T_{+}<T_{-}\}$ with respect to its ``quasi-time-reversal''
trajectory $r\omega$ in $\{T_{+}>T_{-}\}$ is invariable, which is
surprisingly always equal to the constant
$\gamma=\frac{k_1k_2k_3}{k_{-1}k_{-2}k_{-3}}$.

Rigorous proof needs to be expressed in the language of measure
theory, especially applying the Radon-Nikodym derivative similar to
 \cite[Lemma 2.2.7]{JQQ}, so more details is omitted here.

Furthermore, the map $r$ is a one-to-one correspondence between the
trajectory sets $\{T_{+}<T_{-}\}$ and $\{T_{+}>T_{-}\}$. More
particularly, for each $t\geq 0$, the map $r$ is also actually a
one-to-one correspondence between the trajectory sets
$\{T_{+}=t<T_{-}\}$ and $\{T_{+}>T_{-}=t\}$.

Therefore, for each $t\geq 0$,
$$P_{\{E\}}(T_{+}=t,T_{+}<T_{-})=\gamma
P_{\{E\}}(T_{-}=t,T_{-}<T_{+}),$$ and
$$p^{+}=P_{\{E\}}(T_{+}<T_{-})=\gamma
P_{\{E\}}(T_{-}<T_{+})=\gamma p^{-},$$ which has already been proved
in the above section.

Denote the conditional probability density of $T_{+}$ given that
$\{T_{+}<T_{-}\}$ as $\Theta_{+}(t)=P_{\{E\}}(T_{+}=t|T_{+}<T_{-})$,
and the conditional probability density of $T_{-}$ given that
$\{T_{-}<T_{+}\}$ as $\Theta_{-}(t)=P_{\{E\}}(T_{-}=t|T_{-}<T_{+})$.
Hence,

\begin{eqnarray}
\Theta_{+}(t)&=&P_{\{E\}}(T_{+}=t|T_{+}<T_{-})=\frac{P_{\{E\}}(T_{+}=t,T_{+}<T_{-})}{P_{\{E\}}(T_{+}<T_{-})}\nonumber\\
&=&\frac{\gamma P_{\{E\}}(T_{-}=t,T_{-}<T_{+})}{\gamma
P_{\{E\}}(T_{-}<T_{+})}=P_{\{E\}}(T_{-}=t|T_{-}<T_{+})=\Theta_{-}(t),~\forall
t.
\end{eqnarray}

And also denote the probability density of $T$ as
$\Theta(t)=P_{\{E\}}(T=t)$, so

$$\Theta(t)=\Theta_{+}(t)p^{+}+\Theta_{-}(t)p^{-}=\Theta_{+}(t)=\Theta_{-}(t).$$

It consequently follows a very important corollary that the
distribution of waiting cycle time $T$ is \emph{independent} of
whether the enzyme $E$ completes a forward cycle or a backward
cycle, although the probability of these two cycles might be rather
different, i.e.

$$P_{\{E\}}(T=t,T_{+}<T_{-})=P(T_{+}=t,T_{+}<T_{-})=\Theta_{+}(t)p^{+}=\Theta(t)p^{+},$$
and
$$P_{\{E\}}(T=t,T_{+}>T_{-})=P(T_{-}=t,T_{+}<T_{-})=\Theta_{-}(t)p^{-}=\Theta(t)p^{-}.$$

Furthermore, we have

$$\langle T_{+},
T_{+}<T_{-} \rangle=p^{+}\langle T \rangle,$$
$$\langle T_{-},
T_{-}<T_{+} \rangle=p^{-}\langle T \rangle,$$ and
$$\langle T_{+}|T_{+}<T_{-} \rangle= \langle T_{-}|T_{-}<T_{+} \rangle=\langle T \rangle,$$
which means even in the far from equilibrium case ($\gamma>>1$), the
dwell times for each forward cycle or each backward cycle are
identical although their frequencies may be rather different
($p^{+}>>p^{-}$).

At the end of this section, we will present an interesting corollary
about the entropy production rate $e_p$. Due to the classical result
of entropy production rate in a general mesoscopic model of
biochemical kinetic diagrams \cite{Hi77,Hi95,JQQ}, one has

$$e_p=(J^{ss}_{+}-J^{ss}_{-})\log\gamma,$$
where $\log\gamma=\log\frac{J^{ss}_{+}}{J^{ss}_{-}}$ is the entropy
production rate of the cycle $E\rightarrow ES\rightarrow
EP\rightarrow E$, and $J^{ss}_{+}-J^{ss}_{-}$ is its net cycle flux.

Applying the above results to waiting cycle times, $e_p$ can be
expressed as
\begin{eqnarray}
e_p&=&(\frac{1}{\langle T_{+}\rangle }-\frac{1}{\langle T_{-}\rangle})\log\gamma\nonumber\\
&=&(\frac{p^{+}}{\langle T\rangle }-\frac{p^{-}}{\langle T\rangle})\log\gamma\nonumber\\
&=&(p^{+}-p^{-})avepr,
\end{eqnarray}
where $avepr=\frac{1}{\langle T\rangle}\log\gamma=\frac{1}{\langle
T\rangle}\log\frac{J^{ss}_{+}}{J^{ss}_{-}}=\frac{1}{\langle
T\rangle}\log\frac{\langle T_{-}\rangle }{\langle
T_{+}\rangle}=\frac{1}{\langle T\rangle}\log\frac{p^{+}}{p^{-}}$ is
regarded as the time-averaged entropy production rate of the cycle
$E\rightarrow ES\rightarrow EP\rightarrow E$.

Finally, it should be emphasized that this entropy production rate
can also be measured by
$(\nu_{+}(t)-\nu_{-}(t))\log\frac{\nu_{+}(t)}{\nu_{-}(t)}$ when the
time $t$ is large in the single-molecule experiment, recalling that
$J^{ss}_{+}$ and $J^{ss}_{-}$ can be approximated by $\nu_{+}(t)$
and $\nu_{-}(t)$ respectively.

\section{Extending to the $n$-step cycle}

In the previous section, most of the results are obtained through
solving a number of master equations similar to
(\ref{Mean-cycletime}). But now we claim that the same results can
be extended to the $n$-step cycle \cite{GeJ}, according the
elementary renewal theorem in probability theory  \cite[Sec. 3.4,
Theorem 4.1,4.2]{Dur} and general circulation theory of Markov
chains \cite[Chapter 1,2]{JQQ}, which has already been derived for
more than two decades. But the key method is also the same
``quasi-time-reversal'' mapping $r$ introduced in the previous
section.

Below is the summation of the main results in the $n$-step cycle,
which is quite similar to the $3$-step cycle.

\subsection{Cycle flux and NESS}

We consider a n-step mechanism of the Michaelis-Menten kinetics in
which the conversion of $S$ into $P$ in the catalytic site of the
enzyme is represented as a process separate from release of $P$ from
the enzyme.

\begin{equation}\label{Mech-n}
E+S\overset{k_1^0}{\underset{k_{-1}}{\rightleftharpoons}}ES(=ES_1)\overset{k_2}{\underset{k_{-2}}{\rightleftharpoons}}ES_2\cdots
\overset{k_{n-1}}{\underset{k_{-(n-1)}}{\rightleftharpoons}}EP(=ES_{n-1})\overset{k_n}{\underset{k_{-n}^0}{\rightleftharpoons}}E+P,
\end{equation}
in which $k_1^0$ and $k^0_{-3}$ are second-order rate constants.

If there is only one enzyme molecule, then from the enzyme
perspective, the kinetics are stochastic and cyclic, with the
pseudo-first-order rate constants $k_1=k_1^0c_S$ and
$k_{-n}=k_{-n}^0c_P$ where $c_S$ and $c_P$ are the sustained
concentrations of substrate $S$ and $P$ in the steady state.

This system is at chemical equilibrium if and only if
\begin{equation}
\frac{k_1k_2k_3\cdots k_n}{k_{-1}k_{-2}k_{-3}\cdots k_{-n}}=1.
\end{equation}

In the nonequilibrium case,
\begin{equation}
\frac{k_1k_2k_3\cdots k_n}{k_{-1}k_{-2}k_{-3}\cdots
k_{-n}}=\gamma\neq 1,
\end{equation}
and $\triangle\mu=k_BT\ln \gamma$ is well known as the cellular
phosphorylation potential.

Denote a $n$-dimensional matrix $Q=\{q_{ij}\}_{n\times n}$ in which
$q_{i,i+1}=k_i$, $q_{i,i-1}=k_{-(i-1)}$, $i=2,\cdots,n-1$,
$q_{1,2}=k_1, q_{n,1}=k_n, q_{1,n}=k_{-n}, q_{n,n-1}=k_{-(n-1)}$,
and others are all zero. And let $D(H)$ be the determinant of $Q$
with rows and columns indexed by the index set $H$.

Then according to \cite[Theorem 2.1.2]{JQQ}, the enzyme turnover
rate of $S\rightarrow P$, which corresponds to the net flux of the
$n$-step cycle, can be expressed as
\begin{equation}
J^{ss}=\frac{k_1k_2k_3\cdots k_n-k_{-1}k_{-2}k_{-3}\cdots
k_{-n}}{\sum_{i=1,2,\cdots,n}D(\{i\}^c)}=J^{ss}_{+}-J^{ss}_{-},
\end{equation}
where
$$J^{ss}_{+}=\frac{k_1k_2k_3\cdots k_n}{\sum_{i=1,2,\cdots,n}D(\{i\}^c)},$$
is the forward cycle flux, and
$$J^{ss}_{-}=\frac{k_{-1}k_{-2}k_{-3}\cdots
k_{-n}}{\sum_{i=1,2,\cdots,n}D(\{i\}^c)},$$ is the backward cycle
flux.

It should be noticed that the expression of
$\sum_{i=1,2,\cdots,n}D(\{i\}^c)$ is equivalent to the King-Altman
method \cite[Chapter 4]{CB} but more  general and applicable.
Furthermore, it can be easy simulated by mathematical softwares,
such as Matlab and Mathematica.

The net cycle flux can also be expressed as the Michaelis-Menten
steady-state flux of (\ref{Mech-n}), i.e.
$$v=\frac{k_Sc_S-k_Pc_P}{1+\frac{c_S}{K_{mS}}+\frac{c_P}{K_{mP}}},$$
where the definitions of $k_S$, $k_P$, $K_{mS}$, and $K_{mP}$ are
much more complicated than the $3$-step cycle. That is just Eq.
(2.46) in \cite{CB}.

Also similar to the 3-step cycle, $J^{ss}_{+}$ and $J^{ss}_{-}$ can
be rigorously proved to be the averaged numbers of the forward and
backward cycles per time respectively due to ergodic theory
\cite[Theorem 2.1.2]{JQQ}, i.e.

\begin{eqnarray}\label{erg-cycle}
J^{ss}&=&\lim_{t\rightarrow\infty} \frac{1}{t} \nu(t)
,\nonumber\\
J_{+}^{ss}&=&\lim_{t\rightarrow\infty} \frac{1}{t} \nu_{+}(t)
,\nonumber\\
J_{-}^{ss}&=&\lim_{t\rightarrow\infty} \frac{1}{t} \nu_{-}(t),
\end{eqnarray}
where $\nu_{+}(t)$ and $\nu_{-}(t)$ are the number of occurrences of
forward and backward cycles up to time $t$, and
$\nu(t)=\nu_{+}(t)-\nu_{-}(t)$.

\subsection{Mean waiting cycle times}

Starting from the free enzyme state $E$, three kinds of waiting
cycle times can be defined.  $T$ represents the waiting time for the
occurrence of a forward or a backward cycle,  $T_{+}$ represents the
waiting time for the occurrence of a forward cycle, and  $T_{-}$
represents the waiting time for the occurrence of a backward cycle
respectively.

According to the elementary renewal theorem \cite[Sec.3.4, Theorem
4.1,4.2]{Dur},
$$\langle T \rangle=\lim_{t\rightarrow\infty}\frac{1}{\nu_{+}(t)+\nu_{-}(t)}=\frac{1}{J^{ss}_{+}+J^{ss}_{-}}.$$

Similarly,
$$\langle T_{+} \rangle=\lim_{t\rightarrow\infty}\frac{1}{\nu_{+}(t)}=\frac{1}{J^{ss}_{+}},$$
and
$$\langle T_{-} \rangle=\lim_{t\rightarrow\infty}\frac{1}{\nu_{-}(t)}=\frac{1}{J^{ss}_{-}}.$$

\subsection{²½½ø¸ÅÂÊ}

The stepping probabilities $p^{+}(t)$ and $p^{-}(t)$ up to time $t$
are just the fractions of $\nu_{+}(t)$ and $\nu_{-}(t)$ from the
statistical point of view in experiments, i.e.

$$p^{+}(t)=\frac{\nu_{+}(t)}{\nu_{+}(t)+\nu_{-}(t)},~~p^{-}(t)=\frac{\nu_{-}(t)}{\nu_{+}(t)+\nu_{-}(t)}.$$

According to Eq.\ref{erg-cycle}, one can get the eventual stepping
probability
\begin{eqnarray}
p^{+}\stackrel{def}{=}\lim_{t\rightarrow\infty}p^{+}(t)=\frac{J^{ss}_{+}}{J^{ss}_{+}+J^{ss}_{-}},\nonumber\\
p^{-}\stackrel{def}{=}\lim_{t\rightarrow\infty}p^{-}(t)=\frac{J^{ss}_{-}}{J^{ss}_{+}+J^{ss}_{-}}.
\end{eqnarray}

Moreover, due to Eq.(\ref{ratio-stepping}) in the next subsection,
one can prove that the forward stepping probability could also be
defined as $p^{+}\stackrel{def}{=}P_{\{E\}}(T_{+}<T_{-})$, which
means the probability that the particle first completes a forward
cycle before a backward one, starting from the initial state
$\{E\}$. Similarly, the backward stepping probability could be
defined as $p^{-}\stackrel{def}{=}P_{\{E\}}(T_{-}<T_{+})$, too.

Consequently,
$$p^{+}=\frac{J^{ss}_{+}}{J^{ss}_{+}+J^{ss}_{-}}=\frac{\langle T \rangle}{\langle T_{+} \rangle},$$
$$p^{-}=\frac{J^{ss}_{-}}{J^{ss}_{+}+J^{ss}_{-}}=\frac{\langle T \rangle}{\langle T_{-} \rangle},$$
and
$$\triangle\mu=k_BT\log\frac{p^{+}}{p^{-}}=k_BT\log\frac{J^{ss}_{+}}{J^{ss}_{-}}=k_BT\log\frac{\langle T_{-} \rangle}{\langle T_{+} \rangle}.$$

\subsection{Generalized Haldane equality}

Also introduce the same one-to-one mapping $r$ for the trajectory of
the $n$-step kinetic, which belongs to the event $\{T_{+}<T_{-}\}$,
mapped to its ``quasi-time-reversal'' one.

And recall that the number of each forward step in the original
trajectory $\omega$ belonging to $\{T_{+}<T_{-}\}$ of the $n$-step
model is one more than that in its corresponding quasi-time-reversal
trajectory $r\omega$ belonging to $\{T_{+}>T_{-}\}$, and on the
contrary the number of each backward steps in $\omega$ is one less
than that in $r\omega$. And the dwell times of the trajectory
$\omega$ and $r\omega$ are mapped quite well such that the
difference between $\omega$ and $r\omega$ are only exhibited in
their sequences of states.

Consequently, the most important observation is that the ratio of
the probability density of every trajectory $\omega$ in
$\{T_{+}<T_{-}\}$ with respect to its ``quasi-time-reversal''
$r\omega$ in $\{T_{+}>T_{-}\}$ is invariable, which is surprisingly
always equal to the constant $\gamma=\frac{k_1k_2k_3\cdots
k_n}{k_{-1}k_{-2}k_{-3}\cdots k_{-n}}$.

On the other hand, the trajectory map $r$ is a one-to-one
correspondence between the trajectory sets $\{T_{+}<T_{-}\}$ and
$\{T_{+}>T_{-}\}$. More particularly, for each $t\geq 0$, the map
$r$ is also actually a one-to-one correspondence between the
trajectory sets $\{T_{+}=t<T_{-}\}$ and $\{T_{+}>T_{-}=t\}$.

Therefore,
\begin{equation}\label{ratio-stepping}
p^{+}=\gamma p^{-},
\end{equation}
which can be used to prove the results of stepping probabilities in
the previous subsection. Then one arrives at the generalized Haldane
equality in the version of distribution,
\begin{eqnarray}\label{Gen-Hal-dist}
\Theta_{+}(t)&=&P_{\{E\}}(T_{+}=t|T_{+}<T_{-})=P_{\{E\}}(T_{-}=t|T_{-}<T_{+})=\Theta_{-}(t),~\forall
t\geq 0,
\end{eqnarray}
and
$$\Theta(t)=P_{\{E\}}(T=t)=\Theta_{+}(t)p^{+}+\Theta_{-}(t)p^{-}=\Theta_{+}(t)=\Theta_{-}(t).$$

It consequently follows a very important corollary that the
distribution of waiting cycle time $T$ is \emph{independent} of
whether the enzyme $E$ completes a forward cycle or a backward
cycle, although the probability of these two cycles might be rather
different.

Hence, the generalized Haldane equality in the version of
conditional expectation is
\begin{equation}\label{Gen-Haldane}
\langle T_{+}|T_{+}<T_{-} \rangle= \langle T_{-}|T_{-}<T_{+}
\rangle=\langle T \rangle,
\end{equation}
which means even in the far from equilibrium case, the dwell times
for each forward cycle or each backward cycle are the same although
their frequencies might be rather different.

Similar to the previous section, we present an interesting corollary
about the entropy production rate
$e_p=(J^{ss}_{+}-J^{ss}_{-})\log\gamma,$ and
\begin{eqnarray}
e_p&=&(\frac{1}{\langle T_{+}\rangle }-\frac{1}{\langle T_{-}\rangle})\log\gamma\nonumber\\
&=&(\frac{p^{+}}{\langle T\rangle }-\frac{p^{-}}{\langle T\rangle})\log\gamma\nonumber\\
&=&(p^{+}-p^{-})avepr,
\end{eqnarray}
where $avepr=\frac{1}{\langle T\rangle }\log\gamma=\frac{1}{\langle
T\rangle}\log\frac{J^{ss}_{+}}{J^{ss}_{-}}=\frac{1}{\langle
T\rangle}\log\frac{\langle T_{-}\rangle }{\langle
T_{+}\rangle}=\frac{1}{\langle T\rangle}\log\frac{p^{+}}{p^{-}}$ is
regarded as the time-averaged entropy production rate of the
$n$-step cycle.

Furthermore, the following statements are equivalent to each
other:\\
1) This $n$-step system is at chemical equilibrium;\\
2) The cellular phosphorylation potential $\triangle\mu$ vanishes, i.e. $\gamma=1$; \\
3) The forward and backward cycle fluxes are identical, i.e. $J^{ss}_{+}=J^{ss}_{-}$;\\
4) The waiting times for the forward and backward cycles are
identical, i.e. $\langle T_{+}\rangle=\langle
T_{-}\rangle$;\\
5) The eventual stepping probabilities of the forward and backward
cycles $p^{+}=p^{-}$.

Finally, it should be emphasized that this entropy production rate
$e_p$ can also be measured by
$(\nu_{+}(t)-\nu_{-}(t))\log\frac{\nu_{+}(t)}{\nu_{-}(t)}$ when the
time $t$ is large in the single-molecule experiment, recalling that
$J^{ss}_{+}$ and $J^{ss}_{-}$ can be approximated by $\nu_{+}(t)$
and $\nu_{-}(t)$ respectively.

\section{Experimental and theoretically based evidence}

Several results proved in the present article have been observed and
discovered in the recently reported single-molecule experiment
\cite{CC} of kinesin, which is one of the most important molecular
motor proteins. The time trajectories of single kinesin molecules
have been measured for different external forces and for different
ATP concentrations, recording the number of forward and backward
steps ($\nu_{+}(t),~\nu_{-}(t)$) and the stepping probabilities
($p^{+}(t),~p^{-}(t)$), which are also called fractions of forward
and backward steps. They investigated the physical mechanism of the
kinesin step, and found that both forward and backward 8-nm steps
occur on the microsecond timescale without mechanical substeps on
this timescale. It was also shown in \cite[Fig. 3]{CC} that the time
constants for the ensemble averaged forward and backward steps are
very similar, which is just the generalized Haldane equality ) in
the version of conditional expectation (\ref{Gen-Haldane}). However,
this interesting observation has not attached enough importance to
in \cite{CC} and there was no theoretical analysis either.

Almost published at the same time, Kolomeisky, A.B., et.al.
\cite{KSP} put forward a discrete-time biased random walk model,
applying the first-passage-time method \cite{PC} and splitting
probability theory \cite[Chap. XI]{Van}, to provide explicit
expressions for the fractions of forward and backward steps and
dissociations, and most important to conclude that the mean dwell
times to move forward, backward, or irreversible detach are equal to
each other, independent of ATP concentrations or external forces
\cite[Fig.3]{KSP}. The concept of splitting probability in
\cite{KSP,Van} is the same as the stepping probability in the
present paper, and the mean dwell times to move forward and backward
in the discrete-time model in \cite{KSP} just correspond to the
conditional expectation of $T_{+}$ given $T_{+}<T_{-}$ and the
conditional expectation of $T_{-}$ given $T_{-}<T_{+}$ in the
continuous-time model of the present paper respectively, hence what
they conclude was also the generalized Haldane equality in the
version of conditional expectation (Eq. \ref{Gen-Haldane}). They
also claimed that these forward and backward dwell times should be
independent of what direction the motor protein will go in the next
step, although the probability of these steps might be rather
different. Nonetheless, they didn't notice the forward and backward
fluxes and generalized Haldane equality in the version of
distribution (\ref{Gen-Hal-dist}), and it is a pity that they didn't
consider the continuous-time case which is actually more difficult
to prove.

Meanwhile, Kou, S.C., et.al. \cite{KCMEX} summarized their
theoretical understanding of single-molecule kinetics, and focused
on the conditions under which a single-molecule Michaelis-Menten
equation for the reciprocal of the mean stochastic waiting time $T$
for individual turnovers (\cite[Eqs. 14, 26, 30, 34]{KCMEX}). As
have been mentioned in the introduction, the model they built is the
simplest irreversible Michaelis-Menten mechanism, and the state
space of their stochastic model (Markov chain) actually only
contains two states ($E$ and $ES$), which is just why they can just
only use the ordinary differential equations to get the explicit
distribution function $f(t)$ of the waiting time or its
approximations, avoiding to apply the strong Markov property which
is the basic method to compute the first-passage-time problems in
stochastic processes. So their method can not be generalized to more
complicated cases, and generally speaking, the explicit distribution
function $f(t)$ can rarely be analytically obtained in a multi-state
and reversible stochastic model. Moreover, they didn't notice the
existence of forward and backward waiting cycle times $T_{+}$ and
$T_{-}$, and neglected many insightful observations.

Afterwards, Qian, H. and Xie, X.S.\cite{QX} studied a semi-Markov
model of single-enzyme turnover in nonequilibrium steady states with
sustained concentrations of substrates and products, since in some
sense, the general result for \emph{any} enzyme kinetics, if there
is only one free enzyme state, can be mapped to a semi-Markov
process. Then they gave a brief proof to the generalized Haldane
equality in the version of distribution (\ref{Gen-Hal-dist}) and
also expressed the nonzero chemical driving force $\triangle\mu$ as
$k_BT\log\frac{\nu_{+}(t)}{\nu_{-}(t)}$. Hence a part of our results
in the present paper are a special case of the rigorous semi-Markov
result that Wang and Qian have obtained by the elementary renewal
theorem \cite{QW1,QW2}. But they didn't explicitly  distinguish  the
\emph{conditional} waiting cycle time and the \emph{absolute}
waiting cycle times, and may cause some ambiguities.

\begin{figure}[h]
\centerline{\includegraphics[width=3in,height=4in,angle=270]{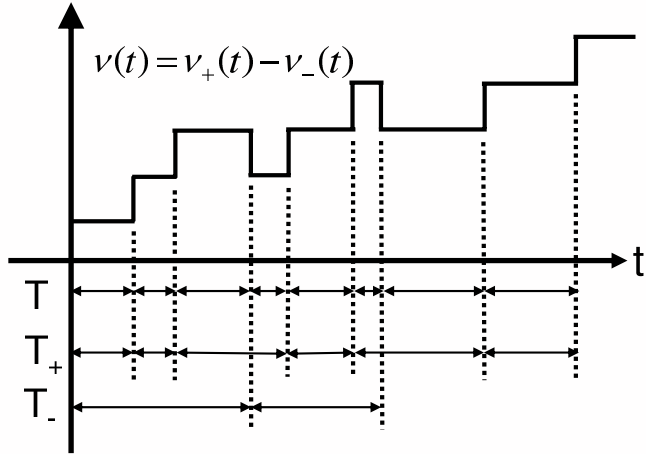}}
\caption[fig4]{The solid line illustrates ideal data on
single-enzyme cycling as a function of time, with distinguished data
analysis of $T$, $T_{+}$ and $T_{-}$. See text for details.}
\label{fig4}
\end{figure}

At the end of this section, what should be paid more attention to in
the statistical data analysis of the experiment is to distinguish
the data of waiting cycle times $T$, $T_{+}$ and $T_{-}$. In
reference \cite{CC}, they only recorded the data of $T$, and divided
them into two classes according to whether have completed a forward
or a backward cycle. Consequently they found the mean values of the
data in these two classes are very similar. But they haven't
realized that the distributions of the data in these two classes
should also be very similar, and actually they are data of the
conditional waiting cycle times rather than the absolute waiting
cycle times $T$, $T_{+}$ and $T_{-}$, because in the nonequilibrium
case, $\langle T_{+}\rangle$ and $\langle T_{-}\rangle$ can not be
identical and must be both greater than $\langle T\rangle$. See Fig.
\ref{fig4} for an illustrative example.

According to the strong Markov property in the theory of stochastic
processes, we claim that the data of $T_{+}$ and $T_{-}$ can also be
obtained from the same trajectory recorded in the experiment (Fig.
\ref{fig4}), and we believe that other main results in the present
paper can also be discovered from the same experiment data,
especially the relationships between the cycle fluxes, mean cycle
times and eventual stepping probabilities, which we have summarized
in the previous section.

\section{Discussion}

Deterministic, nonlinear mathematical models usually based on the
law of mass action have been traditionally used for modelling
biological systems \cite{FMWT,Mu}, while nowadays stochastic
fluctuations observed in most living organisms, such as evidences in
the single-molecule approach \cite{LXX,BWXW}, have changed the way
biophysical or biochemical problems are presented and have been
recognized as a major important effect in cell biology. Stochastic
models in biochemistry have already provided important insights and
quantitative characterizations of a wide range of biochemical
systems \cite{Hi77,Mc,Van,Fox1,Fox2,QH1,Zhou}.

In single-molecule experiments, the microscopic motion of an enzyme
molecule undergoes rapid thermal fluctuation due to its incessant
collisions with the solvent molecules, and therefore, the data
obtained are inevitably stochastic \cite{Xie-rev1}. The main results
in the present paper are actually based on one type of measurements
in single-molecule enzymology, which records the stochastic
conformational dynamics of an enzyme turnover (called
``\emph{trajectories}'') \cite{LXX}. From the perspective of the
stochastic process, a trajectory is a stationary stochastic process
which can be analyzed by statistical methods. And moreover based on
the ergidic theory, which is an elementary law in both statistical
physics and the mathematical theory of stochastic process, an
arbitrary single trajectory surprisingly contains all the
information of the stochastic system.

However, it is often thought that the noise added to the biological
models only provides moderate refinements to the behaviors otherwise
predicted by the classical deterministic system description, while
in the present paper, it is quite obvious that the main problems
discussed here are \emph{impossible} even to be put forward in a
deterministic model. So it may be necessary to reconstruct the main
biological theory based on the stochastic models in order to explain
the experiment results of single molecule tracking.

For instance, applying the statistical methods to analysis the
recorded single trajectory, experimentalists can not only directly
measure the distribution of the waiting cycle (turnover) times $T$,
$T_{+}$ and $T_{-}$, but also the probability cycle fluxes $J^{ss}$,
$J^{ss}_{+}$ and $J^{ss}_{-}$ widely used in this paper£¬which can
be approximated by the time-averaged number of occurrences of the
cycle, according to the general ergodic theory of cycle fluxes
\cite[Theorem 2.1.2]{JQQ} and elementary renewal theorem \cite[Sec.
3.4, Theorem 4.1,4.2]{Dur}. Furthermore, it is important to notice
that the distributions of single-molecule properties and the
probability cycle fluxes can only be presented and measured through
single-molecule experiments rather than the ensemble average.

In conclusion, the single-molecule enzymology is still in its early
ages, and in my personal opinion, the generalized Haldane equality
as well as the relationship between mean waiting cycle times and
cycle fluxes may be the first interesting discovery in this active
field, which could be applied to instruct the data analysis of
single-molecule trajectories. In the future, we believe that more
and more important phenomenon and theory in traditional enzymology,
such as inhibition and activation, cooperativity and multi-enzyme
systems, will enter into the single-molecule enzymology, which might
stimulate very significant developments both in experiment and
theory.

In addition, from the purely mathematical point of view, we also
believe that it is valuable to further extend similar results to the
theory of much more general Markov chains, though much technical
work remains to be done.

\section*{Acknowledgement}

The authors would like to thank Prof. Hong Qian at University of
Washington for introducing this problem, and Prof. Min Qian, Prof.
Maozheng Guo, Prof. Daquan Jiang and Yuping Zhang in Peking
University for helpful discussion in our seminar. This work is
partly supported by the NSFC (Nos. 10701004, 10531070 and 10625101)
and 973 Program 2006CB805900.


\begin{thebibliography}{99}
\bibitem{GeJ} Mathematical rigorous proof based on measure theory is being prepared.
\end{thebibliography}
\end{document}